# Understanding bias in the introduction of variation as an evolutionary cause[1]

Arlin Stoltzfus (arlin@umd.edu)

Fellow, IBBR, 9600 Gudelsky Dr., Rockville, MD, 20850

Research Biologist, NIST, 100 Bureau Dr., Gaithersburg, MD 20878

## Abstract

Our understanding of evolution is shaped strongly by how we conceive of its fundamental causes. In the original Modern Synthesis, evolution was defined as a process of shifting the frequencies of available alleles at many loci affecting a trait under selection. Events of mutation that introduce novelty were not considered evolutionary causes, but proximate causes acting at the wrong level. Today it is clear that long-term evolutionary dynamics depend on the dynamics of mutational introduction. Yet, the implications of this dependency remain unfamiliar, and have not yet penetrated into high-level debates over evolutionary theory. Modeling the influence of biases in the introduction process reveals behavior previously unimagined, as well as behavior previously considered impossible. Quantitative biases in the introduction of variation can impose biases on the outcome of evolution without requiring high mutation rates or neutral evolution. Mutation-biased adaptation, a possibility not previously imagined, has been observed among diverse taxa. Directional trends are possible under a sustained bias. Biases that are developmental in origin may have an effect analogous to mutational biases. Structuralist arguments invoking the relative accessibility of forms in state-space can be understood as references to the role of biases in the introduction of variation. That is, the characteristic concerns of molecular evolution, evo-devo and structuralism can be interpreted to implicate a kind of causation absent from the original Modern Synthesis.

---

> Most evolutionary geneticists would agree that the major problems of the field have been solved. We understand both the nature of the mutational processes that generate novel genetic variants and the populational processes which cause them to change in frequency over time— most importantly, natural selection and random genetic drift, respectively . . . we will never again come up with concepts as fundamental as those formulated by the 'founding fathers' of population genetics (Charlesworth 1996)

# Understanding biases in the introduction of variation

## *Climbing Mount Probable*

Imagine, as an analogy for evolution, a climber operating on a rugged mountain landscape (Fig. 1A). A human climber would scout a path to the highest peak and plan accordingly, but an analogy for evolution must disallow foresight and planning, therefore let us imagine a blind robotic climber. The climber will move by a two-step proposal-acceptance mechanism. In the "proposal" or "introduction" step, the robotic climber reaches out with one of its limbs to sample a point of leverage, some nearby hand-hold or foot-hold. Each time this happens, there is some probability of a second "acceptance" step, in which the climber commits to the point of leverage, shifting its center of mass.

Biasing the second step, such that relatively higher points of leverage have relatively higher probabilities of acceptance, causes the climber to ascend, resulting in a mechanism, not just for moving, but for climbing.

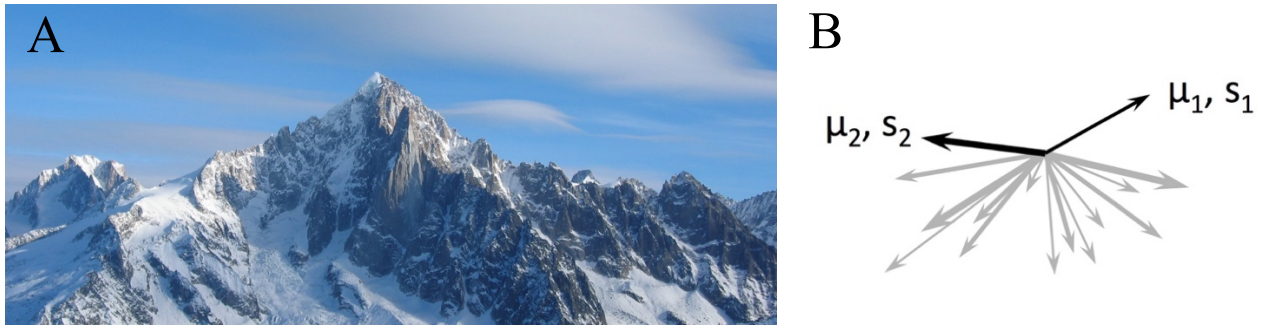

Figure 1. (A) The rugged landscape surrounding Aiguille Verte, Massif du Mont Blanc (photo credit: see Acknowledgements). (B) A case of two possible beneficial mutations. Mutations to alternative points in a state-space are shown as arrows projected onto 2 dimensions, with fitness in the vertical dimension, and with mutation rate represented by line thickness. Two of the mutations lead to more beneficial states, one going to the left, and the other going to the right.

What happens if a bias is imposed on the proposal step? Imagine that the robotic climber (perhaps by virtue of longer or more active limbs on one side) samples more points on the left than on the right during the proposal step. Because the probability of proposal is greater on the left, the joint probability of proposal-and-acceptance is greater (on average), so the trajectory of the climber will be biased, not just upwards, but to the left as well. If the landscape is rough (as in Fig. A), the climber will tend to get stuck on a local peak that is upwards and to the left of its starting point. If the landscape is a smooth hemisphere, the climber will spiral up and to the left, ascending to the single summit.

## *A population-genetic model*

The metaphor of "climbing mount probable" helps us to conceptualize the influence of a bias in the introduction of novelty in evolution— unless, of course, the metaphor is misleading. To find out, we must ask whether we can make it work in terms of population genetics.

In attempting to establish any such principle, one begins with the simplest case. In the simplest case, the climber has access to only two different moves, one going up and to the left, and the other going up and to the right. If the same path is favored by biases in both the proposal step and in the acceptance step, then that path is obviously favored. The non-trivial question is what happens when the leftward outcome is more strongly favored by the proposal step, and the rightward outcome more strongly favored by acceptance. That is, suppose that we have some set of possible mutations (possible moves) illustrated in Fig. 1B. Each mutation goes from the current state to some other state (e.g., from one DNA sequence to another), and each one is characterized by a mutation rate $u$ and a selection coefficient $s$. At some point in time, there are 2 beneficial mutations (black arrows), with the leftward mutation having a higher mutation rate $u_2 > u_1$, but a lower fitness benefit $s_2 < s_1$.

Let $B$ represent the bias in mutation favoring the leftward move, $B = u_2 / u_1$, and let $K$ represent the bias in selection coefficients favoring the rightward move, $K = s_1 / s_2$. How often will evolution climb to the left, instead of the right? How much will $B$ influence the outcome? How will the outcome depend on $K$, or on the strength of selection?

To simulate evolution with this model, we will start with a pure population that may evolve by mutation and fixation to either the left (type 2) or the right (type 1), considering a range of population size $N$ from 100 to a million. We will vary $B$ from 1 to 1000 by setting $u_2 = 10^{-5}$ per generation, and allowing the other mutation rate to vary; meanwhile, $s_1 = 0.02$ and $s_2 = 0.01$, so that $K = 2$. Under these conditions, the time to fixation varies inversely with $N$, on the order of $10^3$ to $10^5$ generations, because in all but the largest populations, most of the time is spent waiting for a beneficial allele to arise and reach sufficient numbers that stochastic loss becomes unlikely.

The results of simulating the model (hundreds of times for each set of conditions) are shown for a range of population sizes in Fig. 2A, with the bias in outcomes— the ratio of left to right outcomes— as a function of mutation bias $B$. All of the lines are going up: the bias in outcomes increases with the bias in mutation. When a line goes above $y = 1$, the mutationally favored outcome prevails. For $N = 100$ or $N = 1000$, the bias in outcomes is approximately $B / K$ (dashed grey line), the outcome expected from origin-fixation dynamics (explained below). That is, under certain conditions, an exact relationship is expected between mutation bias and the outcome of evolution. For larger population sizes, we depart from the origin-fixation regime, but mutation bias continues to influence the outcome of evolution.

Let us consider the behavior of this model in further detail. What about the strength of selection? Is the behavior above simply a matter of selection not being strong enough? Actually, $s = 0.02$ or $s = 0.01$ is considered to be "strong selection" given a population size of $N = 1000$ or higher. But let us demonstrate this point more clearly. In Fig. 2B below, the bias in outcomes is shown as a function of the higher selection coefficient ($s_1$) over a 200-fold range from $s_1 = 0.001$ to $s_1 = 0.2$.

The lines are flat, indicating that the biasing effect of mutation is not some kind of "opposing forces" contest with selection. We have increased the magnitude of *s*, increasing the force of selection, with no effect. If effect (output) does not increase with force (input), then the "force" analogy fails.

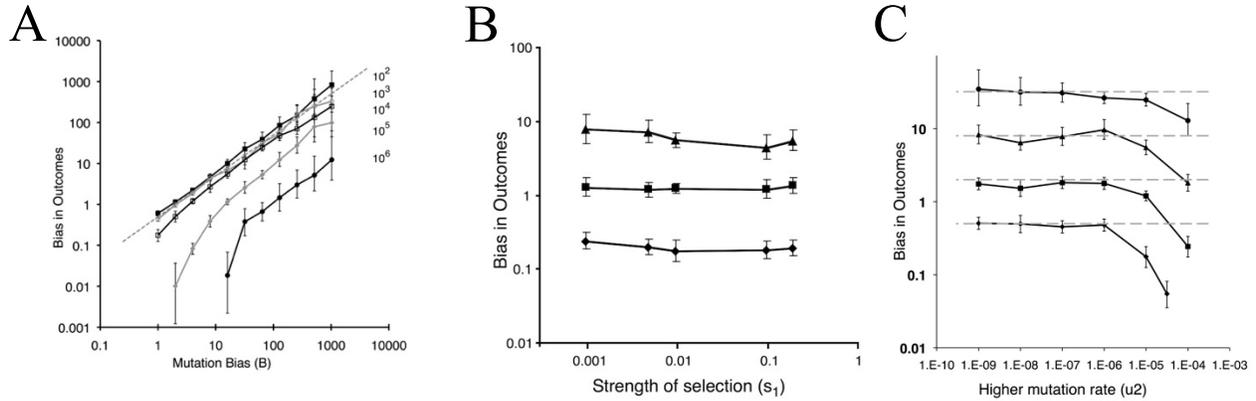

Figure 2. Results of simulations. (A) Effect of population size. For each population size indicated on the right, evolution was simulated over a 1000-fold range of mutation bias, with hundreds of replicates for each set of parameters. (B) Effect of the strength of selection, which here varies 200-fold for *B = 16* (triangles), *B = 4* (squares), and *B = 1* (diamonds), with *N = 1000* and other parameters as before. (C) Effect of the strength of mutation. From top to bottom, the lines here are for *B* values of 64, 16, 4 and 1, where *K = 2*, thus *B / K* (dashed grey lines) is 32, 8, 2 and 0.5. As mutation rate decreases going from right to left, the bias in outcomes approaches *B / K*.

Likewise, we can ask whether the biasing effect is dependent on high mutation rates. Fig. 2C shows the bias in outcomes for 4 different values of *B* as a function of the rate of mutation (the higher rate, $u_2$), which decreases *from right to left* (for *N = 1000* and other conditions the same as above). As the mutation rate decreases, the bias in outcomes converges on *B / K* (dashed grey lines) for each set of conditions. Again, this violates the "force" theory: larger forces must yield larger effects, otherwise the concept is inapplicable.

The behavior above is readily understood from origin-fixation models, which represent evolutionary change as a simple 2-step process of (1) the introduction of a new allele by mutation, followed by (2) its fixation or loss (McCandlish and Stoltzfus 2014). The most familiar version of this formula is *K = 4Nus*, where *2Nu* is the rate of mutational introduction of beneficial alleles with selection coefficient *s*, and *2s* is the probability of fixation for such alleles. For the case of neutral alleles, the probability of fixation is *1 / (2N)*, so the origin-fixation rate of neutral evolution is *K = 2Nu / (2N) = u*.

Origin-fixation models have distinctive implications regarding the effect of biases in the introduction of new alleles (McCandlish and Stoltzfus 2014). For example, consider a model in which a population is currently fixed at allele i and can mutate to a variety of other alleles. Using *u* for a mutation rate and *π* for a probability of fixation (dependent on a selection coefficient *s* and a population size *N*), the odds that the population will next become fixed at allele *j* rather than allele *k* are given by Eqn. 1:

$$\frac{P_{ij}}{P_{ik}} = \frac{2Nu_{ij}\pi(s_{ij}, N)}{2Nu_{ik}\pi(s_{ik}, N)} = \frac{u_{ij}}{u_{ik}} \times \frac{\pi(s_{ij}, N)}{\pi(s_{ik}, N)}$$

This can be described as the product of two factors: a ratio of mutation rates, and a ratio of fixation probabilities. For the case of beneficial alleles, we can use the approximate probability of fixation *2s* (Haldane 1927), and this reduces to Eqn. 2 (Yampolsky and Stoltzfus 2001).

$$\frac{P_{ij}}{P_{ik}} = \frac{u_{ij}}{u_{ik}} \times \frac{s_{ij}}{s_{ik}}$$

The *B* / *K* ratio mentioned above is simply an instance of this relation, that is, $B = u_{ij}/u_{ik}$ and $1/K = s_{ij}/s_{ik}$. Thus, not only does a bias in mutation have a direct, proportional effect on the odds of one outcome relative to another, this effect is neither weaker nor stronger, but exactly the same as, a bias in fixation of the same magnitude. That is, doubling the mutation rate to an allele or doubling its probability of fixation both have the same effect on the odds that it is the next allele to be fixed.

## *The sushi conveyor and the buffet*

The behavior described in the previous section is not universal. In some regimes of population genetics, biases in mutation are completely ineffectual. To understand why, let us consider another analogy, comparing two regimes of population genetics with two styles of self-service restaurant— the buffet and the sushi conveyor (Fig. 3). In the former, we begin with a practically inexhaustible abundance of static choices in full view, and fill our plate with the desired amount of each dish, often choosing many different items; in the latter, we iteratively make a yes-or-no choice of the chef's latest creation as it passes by, typically accepting only a few choices and rejecting the vast majority. In either case, we exercise choice, and we may end up with a satisfying meal.

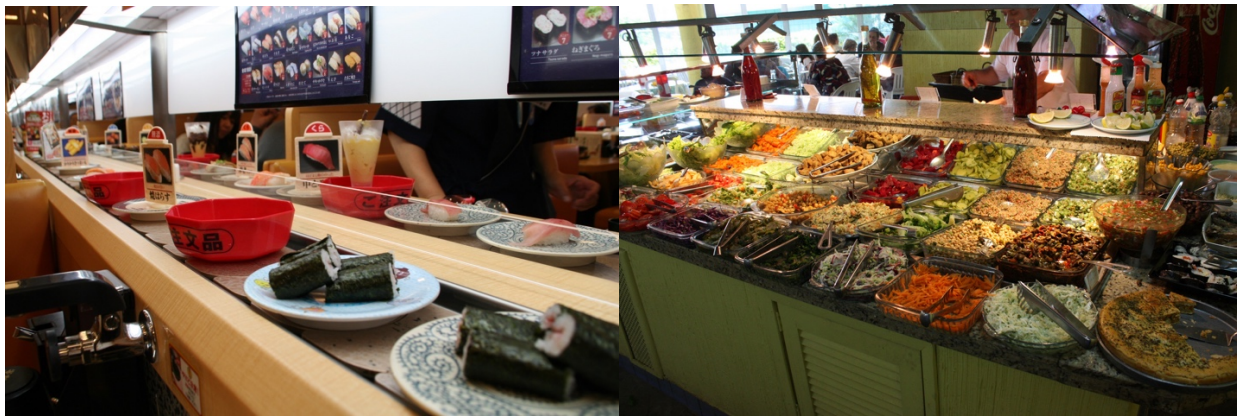

Figure 3. Two styles of self-service restaurant, the sushi conveyor (left) and the buffet (right; photo credits: see Acknowledgements).

The thinking of molecular evolutionists frequently corresponds to the sushi conveyor model, which illustrates a proposal-acceptance process like the one described in "Climbing Mount Probable." We choose (we select), but we don't control what is offered or when: instead, we accept or reject each dish that passes by our table. Though we decide whether each dish is right for us, initiative and creativity belong largely to the chef.

By contrast, the architects of the Modern Synthesis were committed to the buffet view, in the form of a totipotent "gene pool" with sufficient variation to respond to any challenge (addressed in more detail below). Just as the staff who tend the buffet will keep it stocked with a variety of choices sufficient to satisfy every customer, the gene pool is said to *maintain* abundant variation, so that selection never has to wait for a new mutation. Adaptation happens when the customer gets hungry and proceeds to select a platter of food from the abundance of choices, each one ready at hand, choosing just the right amount of each ingredient to make a well balanced meal.

A bias in the choices offered to customers (analogous to a bias in variation) will have different effects in the two regimes. Let us suppose that the buffet has 5 apple pies and 1 cherry pie. This quantitative bias makes no difference. A rational customer who prefers cherry pie is unaffected by relative amounts and will choose a slice of cherry pie every time. The only kinds of biases in supply that are relevant at the buffet are absolute constraints— the complete lack of some possible dish—, that is, the customer who prefers cherry pie will end up with apple pie only if there are no cherry pies.

But at the sushi conveyor, the effect of a bias will be different. Let us suppose that the dishes of sushi on the conveyor show a 5 to 1 ratio of salmon to tuna. Even a customer who would prefer tuna in a side-by-side comparison may eat salmon more often, because a side-by-side comparison simply is not part of the process.

How does this difference-by-analogy map to a difference in regimes of evolutionary genetics? We have already seen sushi-conveyor dynamics (Fig. 2). The question is whether the buffet regime exists, and if so, what are its defining characteristics? The key condition in a hypothetical buffet regime would be that the variants relevant to the outcome of evolution are abundantly present in the initial gene pool.

Therefore, let us consider what is the effect of initial variation in the model presented earlier. The result, shown in Fig. 4, is simple and compelling. Here we have simulated the model exactly as before. The upper line repeats the results (from Fig. 2) of evolution from a pure starting population, and the lower lines represent cases in which both alternative types are present in the initial population at non-zero frequencies (Yampolsky and Stoltzfus 2001). Even when the initial frequencies of the alternatives are just 0.5 %, mutation bias has essentially no effect.

Why is selection so much more powerful in the buffet regime? The 2-fold difference between $s_1 = 0.02$ and $s_2 = 0.01$ corresponds to a 2-fold preference for type 1 in the origin-fixation regime, given that the behavior of selection in this regime follows the probability of fixation, $2s$. In the buffet regime, the impact of exactly the same fitness difference is far greater: if we have 2 alternative alleles already in a population and they have escaped random loss, selection nearly always establishes the more fit alternative, even if it starts out at a lower frequency.

Why is mutation bias so important in the sushi conveyor regime? Fig. 4 shows that a bias in outcomes does not always follow from a bias in mutation: the bias $B$ increases over 3 orders of magnitude, but sometimes this has no effect. If mutation bias is strong but there is no effect on outcomes, i.e., if the putative cause is present but the expected effect is not, this suggests that mutation bias *per se* is not the proper cause. Instead, the bias operating so effectively in the origin-fixation regime is a bias in the *introduction process*. In the buffet regime, there is no bias in the introduction process, because *there is no introduction process*— all relevant alleles are present already.

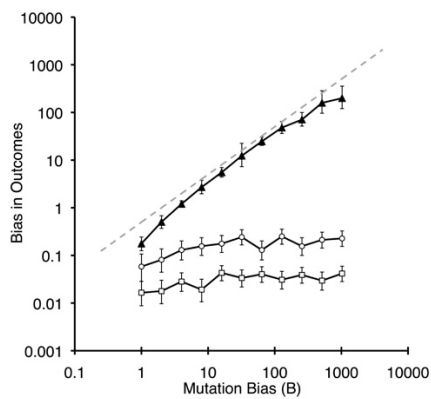

Figure 4. Effect of initial variation in the Yampolsky-Stoltzfus model. The upper series (closed triangles) represents the same results for *N = 1000* shown in Fig. 2A. In the lower two series, the initial population contains both alternative types at a low frequency of either 0.005 (open circles) or 0.01 (open squares).

Note that the above results are from simulations. The behavior of this model does not have obvious mathematical solutions except in two limiting cases. In the limiting case as $u \rightarrow 0$ with $N$ fixed (where $u = u_1 + u_2$), we may consider an absorbing Markov chain whose behavior is readily understood with origin-fixation dynamics, yielding Eqn 2. In the limiting case as $N \rightarrow \infty$ with $u$ fixed, i.e., a deterministic case of shifting frequencies with all types present immediately, we may consider a set of coupled differential equations whose behavior dictates that the fittest type asymptotically approaches a frequency near 1. This behavior corresponds to the buffet regime. In the intermediate regime where concurrent mutation is possible (as per Weissman et al. 2009, Desai and Fisher 2007), the problem has not been solved, yet the simulation results in the preceding section indicate clearly that biases in mutation are important.

## *Distinctive implications*

The results above reveal a kind of cause-effect relationship that seems very fundamental, but is absent from standard treatments of evolutionary causation. Stated more precisely, it is the cause-effect relationship by which biases in the introduction of variation have a difference-making power characterized by the following statements:

- *It applies when fixations are selective*. Hypotheses in which mutation bias shapes features, though not allowed in the original Modern Synthesis (OMS), quickly became accepted by molecular evolutionists as implications of neutrality, under the assumption that mutation bias can be effective only when selection is absent. By contrast, biases in the introduction of variation do not require neutrality, leading to the novel prediction of mutation-biased adaptation.

- *It poses a directional, quantitative dependence*. In conventional thinking, variation is only a material cause, a source of substance but not form, and so the obvious questions to ask about variation concern the amount of raw material available. By contrast, the cause-effect relationship proposed here identifies quantifiable *directions* of variation that may influence directions of evolution, and proposes idealized conditions under which the effect of a mutational bias of magnitude $B$ is a $B$-fold bias on evolution.
- *It is regime-dependent.* The maximal effect just mentioned is fully realized in the sushi conveyor regime, but is negligible in the buffet regime, where the variants relevant to the outcome of evolution are abundantly present. By contrast, conventional thinking assigns *fixed roles* to mutation and selection, independent of population-genetic regime.
- *It depends on the rareness of mutations*. In this kind of causation, the influence of biases in the generation of variation emerges as $uN \rightarrow 0$. By contrast, conventional arguments based on the force of mutation assume that mutation rates must be large in magnitude for mutational effects to be important in evolution.
- *It establishes a condition of parity with selection*. Under ideal conditions, the effect of a bias in the introduction of new alleles is the same in magnitude as that of a bias in fixation probabilities of equal magnitude. By contrast, conventional thinking treats variation as a different *kind* of cause that is passive and subordinate to selection.
- *It establishes a condition for composition and decomposition of causes*. The proposed causal relationship allows that biases in mutation and biases in fitness effects may, under ideal conditions (Eqn 1), both act like vectors that combine to determine the resultant direction of evolution. This provides a basis for combining or disentangling causes.

Such a cause-effect relationship is not described in the canonical works of the Modern Synthesis. It is not found, to my knowledge, in present-day textbooks of evolution. In fact, as will be described further below, a famous argument from Fisher and Haldane says that biases in variation cannot influence the course of evolution. Various examples could be given in which the *absence* of this cause-effect relationship is conspicuous, e.g., if the authors of a famous "developmental constraints" paper (Maynard Smith et al. 1985) had understood this relationship, surely they would have invoked it to resolve the contentious issue of how developmental biases in variation were supposed to act as evolutionary causes.

## *Some evidence*

In the field of molecular evolution, hypotheses that invoke mutational causes for patterns are common (Stoltzfus and Yampolsky 2009). The effects of mutation bias appear to be pervasive in molecular evolution. Nevertheless, when this is interpreted as an aspect of neutral evolution, it is superficially consistent with conventional "forces" thinking: mutation pressure can be effectual when the opposing pressure of selection is absent.

The distinctive implication of biases in the introduction of variation is that their effectiveness, unlike that of mutation pressure, does not depend on neutrality. Thus, in seeking evidence for this kind of causation, we are particularly interested in cases where beneficial changes are happening. Available cases include instances in which mutation bias affects the outcome of experimental adaptation (Cunningham et al. 1997, Rokyta et al. 2005, Meyer et al. 2012, Couce, Rodríguez-Rojas, and Blázquez 2015, Sackman et al. 2017), the argument of Galen, et al. (2015) for the role of a CpG mutational hotspot in altitude adaptation of Andean house wrens (see also Stoltzfus and McCandlish 2015), and a recent meta-analysis of parallel adaptive changes (Stoltzfus and McCandlish 2017).

Cunningham, et al. (1997) adapted bacteriophage T7 in the presence of the mutagen nitrosoguanidine. Deletions evolved 9 times and nonsense codons 11 times, sometimes with the same change occurring multiple times. All the nonsense mutations were GC-to-AT changes, which is the kind of mutation favored by the mutagen. Couce, et al. (2015) carried out experimental adaptation in *E. coli* with replicate cultures from wild-type, mutH, and mutT parents, the latter two being "mutator" strains with elevated rates of mutation and different mutation spectra. When these strains were subjected to increasing concentrations of the antibiotic cefotaxime, the resulting adaptive changes reflected the differences in mutational spectrum between lines: resistant cultures from the mutT parent tended to adapt by a small set of A:T → C:G transversions, while resistant cultures from the mutH parent tended to adapt by another small set of G:C → A:T and A:T → G:C transitions.

Meyer, et al. (2012) reported changes in the J gene of bacteriophage Lambda in 48 replicate cultures of *E. coli*, half of which had acquired the ability to utilize the OmpF receptor as an attachment site. The complete set of 241 differences (all non-synonymous) from the parental J gene in 48 replicates is shown in Fig. 5. Among these non-synonymous mutations, 22 are found at least twice (asterisks), including 16 transitions observed in adapted strains 181 times, and 6 transversions observed 42 times. Thus the transition:transversion ratio is 16/6 = 2.7 when we count paths (i.e., columns in Fig. 5), and 181/42 = 4.3 when we count events (cells in Fig. 5). This is far above the null expectation of 0.5 under an absence of mutation bias.

To explore the role of mutational biases in parallel adaptation more systematically, Stoltzfus and McCandlish (2017) gathered data for a set of experimental cases such as Meyer, et al. (2012), and for a comparable set of natural cases of parallel adaptation. They used these data to test for an effect of transition:transversion bias, a widespread kind of mutation bias. In the data set of 389 parallel events along 63 paths from experimental studies of evolution, they find a highly significant tendency—from 4-fold to 7-fold in excess of null expectations— for adaptive changes to occur by transition mutations rather than transversion mutations. For the data set of natural cases of parallel adaptation consisting of 231 parallel events along 55 paths, they found a bias of 2-fold to 3-fold over null expectations, which was statistically significant for both paths and events. They conclude that parallel adaptation takes place by nucleotide substitutions that are favored by mutation, noting that the size of this effect is not a small shift, but a substantial effect of 2-fold or more.

This analysis was made possible by prior work that used empirical data on the effects of mutations to show that transitions and transversions do not differ importantly in their distributions of fitness effects (Stoltzfus and Norris 2016). This was important to establish because the lore in molecular evolution held that amino acid changes tend to occur by transitions because they are more conservative in their effects (Keller, Bensasson, and Nichols 2007, Rosenberg, Subramanian, and Kumar 2003, Wakeley 1996). Thus, taking the work of Stoltzfus and McCandlish (2017) at face value, a several-fold effect previously attributed to selection is now best understood as being entirely or almost entirely due to mutation.

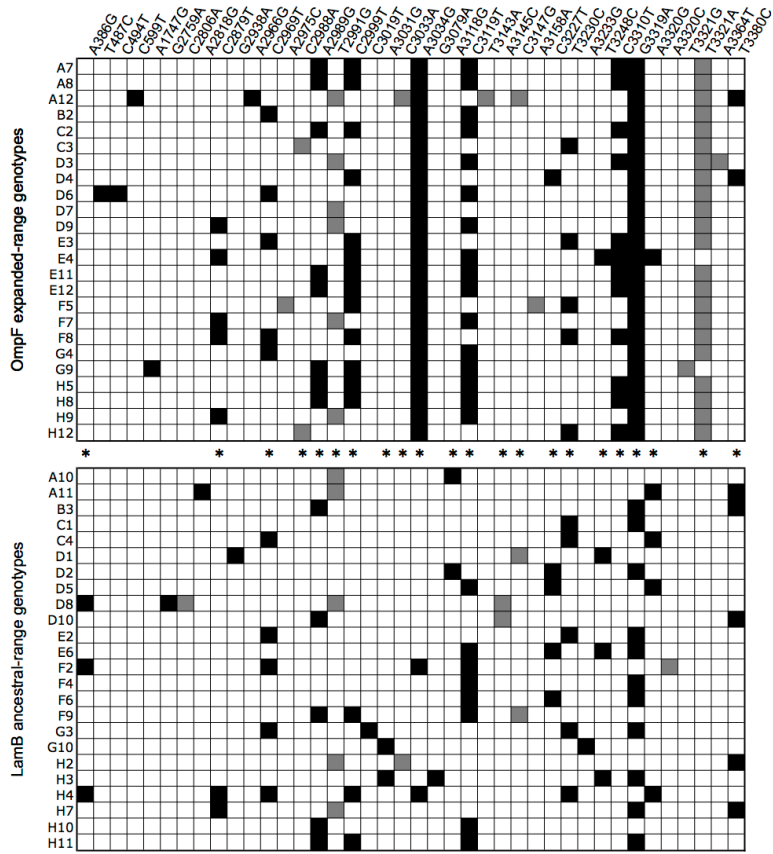

Figure 5. Parallel evolutionary changes (asterisks) in the Lambda tail tip gene (after Fig. 1 of Meyer et al. 2012). Black (transition) and grey (transversion) boxes indicate specific nucleotide mutations (columns) found among 48 replicates (rows).

More generally, I take it that the way to establish that a kind of cause is important in evolution is to show that it has quantitatively large effect-sizes in regard to features that evolutionary biologists consider important or interesting. An effect-size of 2-fold is large. In a data-rich field such as molecular evolution, enormous numbers of papers are published every year based on far smaller effect-sizes. The subjectivity of "important or interesting" cannot be avoided, but may be exploited, as in the above example, by stealing explanatory power from selection: if some evolutionary phenomenon (such as transition:transversion bias) was important enough to elicit a selective explanation, then an alternative mutational explanation is also important.

Note that this evidence does not address all of the distinctive implications outlined above. Specifically, it addresses the claim that an effect of mutation bias is possible under non-neutral conditions. With more data, it will be possible to address a graduated relationship between cause and effect (given that the magnitude of mutation bias varies with taxonomic context), the composition of causes, and the role of population-genetic regime.

# Three larger implications

## *The potential for directional trends*

The model above addresses what theoreticians would call a 1-step adaptive walk. Is it possible for a sustained bias in the introduction of new variation to cause a sustained trend over many steps? To address this question, one must consider a multi-step adaptive walk on some larger landscape.

Precisely this issue is addressed by Stoltzfus (2006), using a model of a protein-coding gene, based on an abstract NK fitness model to represent interactivity of amino acids in a protein. The question addressed in the model is whether a mutational bias toward G and C nucleotides could result in a biased protein composition, even when all of the changes are beneficial. This question is relevant to interpreting observed differences in protein composition that correlate with GC content (Singer and Hickey 2000): organisms with high-GC synonymous sites and inter-genic regions also have proteins enriched for amino acids encoded by high-GC codons (Gly, Ala, Arg, Pro) and depleted for those encoded by low-GC codons (Phe, Tyr, Met, Ile, Asn, Lys). Typically such differences are attributed to mutational biases operating in the context of neutral evolution. However, we can now understand the potential for mutation biases to operate in non-neutral evolution.

The general result of simulating evolution of a protein-coding gene via beneficial changes, subject to biases in the introduction of variants, is to confirm the intuitions of the metaphor of Climbing Mount Probable, showing that it is not misleading in its implications. On a perfectly smooth landscape, the trajectory of evolution is deflected initially by mutation bias, but the ultimate destination— the optimal sequence— is unaffected. On rough landscapes, a mutation-biased composition evolves along a trajectory to a local peak. The rougher the landscape, the shorter the trajectory and the greater the per-step effect of mutation bias. In the model used by Stoltzfus (2006), a modest mutational bias toward GC (or AT), with a magnitude consistent with mutational biases inferred from models of genome composition evolution, can cause a progressive shift in protein composition of a magnitude consistent with observed biases in protein composition.

## *The causal efficacy of developmental biases*

The novel implications of biases in the introduction process apply to some biases that are properly developmental, in the sense of reflecting the mapping of genotypes to phenotypes. The simplest and most complete such mapping is the genetic code, which represents the developmental process of translation, by which an amino acid emerges as the expression of a triplet codon. Like other developmental processes, translation is a complex, multi-step process. Like other developmental processes, there is not a 1:1 mapping of phenotypes to genotypes: most amino acid phenotypes can be encoded in multiple ways; some changes are phenotypically silent. Unlike many other developmental processes, translation results in discrete phenotypes. More distinctly, translation results in discrete phenotypes in such a highly canalized way that the developmental realization of the modal phenotype occurs over 99.9 % of the time, i.e., less than 1 in 1000 amino acids is the wrong one.

Though translation isn't the typical case of development discussed in the evolutionary literature, its differences from the typical cases of discrete phenotypic variants are a matter of degree. Many developmental processes result in discrete phenotypes, e.g., limb development in chordates results in an autopod with a discrete set of digits, with a modal number of 5 for mammals. Cranial development is so strongly canalized that nearly every mammal is born with precisely 1 head and not 2, 3 or more. Yet, rare cases indicate that double-headedness is developmentally possible (typically with duplication of the neck and upper torso), with a chance of occurrence less than that of a translation error.

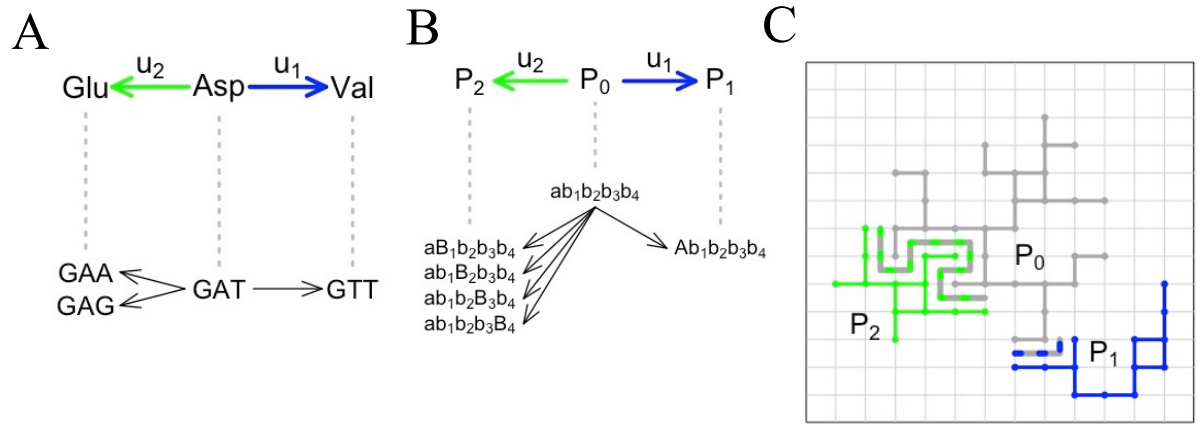

Figure 6. Biases induced by features or processes above the DNA level. (A) A 2-fold bias in mutational paths induced by the genetic code. (B) A 4-fold bias in mutational paths induced by an asymmetric mapping of genotypes to phenotypes. One mutation (a → A) at the A locus leads to $P_1$, and 4 different mutations at loci $B_1$, $B_2$, $B_3$ and $B_4$ lead to $P_2$. (C) Differential accessibility of phenotypic states induced by the structure of connected neutral networks in sequence space (after Fontana 2002). Only 3 networks are shown, each connecting RNA sequences with the same fold (phenotype). Over time, a sequence traverses a network neutrally, and the chance of evolving to a different network (phenotype) depends on its proximity. Because the boundary (thick dashed lines) between $P_0$ and $P_2$ is several times longer than between $P_0$ and $P_1$, $P_0$ is several-fold more likely to evolve to $P_2$.

Consider a single codon encoding an amino acid as an evolving system. The system begins with the Aspartate phenotype, encoded by GAT, as in Fig. 6A. The developmental expression of a genotype is indicated by the dotted grey line. From this starting point, there are twice as many mutational paths to genotypes expressed as the Glutamate phenotype than expressed as Valine. Assuming (for the sake of example) that all individual mutation rates are the same, this means that, in a population of Asp-encoding genotypes, there will be a 2-fold bias in the introduction of alternative Glu phenotypes, relative to Val phenotypes.

The asymmetry described above is analogous to a form of developmental bias invoked repeatedly in the evo-devo literature, by which some phenotypes are more likely to arise by mutation than others (e.g., Alberch and Gale 1985, Emlen 2000). Fig. 6B gives this concept a precise interpretation, in which, given the starting phenotype $P_0$, there are 4 times as many ways to mutate to $P_2$ as to $P_1$. In the language of genetics, the mutants that generate $P_2$ are phenocopies of each other. This kind of bias differs formally from a DNA-level bias in that it cannot be defined in terms of DNA chemistry, but must make reference to development or phenotypes.

The evolutionary impact of such biases can be understood by mapping them to the Yampolsky-Stoltzfus model. We could define $u_2$ to cover mutations from Asp to Glu, whether they are to GAA or GAG, and $u_1$ to cover mutations to Val. Or we could define $u_2$ to cover all 4 mutations from $P_0$ to $P_2$, and $u_1$ to cover mutations to $P_1$. Then, all of the same implications would apply, and in particular, the 2-fold or 4-fold bias will exert a 2-fold or 4-fold bias (respectively) on the outcome of evolution in the origin-fixation regime.

## *An interpretation of structuralism*

Classic thinking in evolutionary biology includes a preference (often implicit) for explanations that are externalist, functionalist, and mechanistic: the explanations have to do with external rather than internal causes; they explain form by relating it to function; and they are seen to rest on Aristotelean material and efficient causes rather than formal causes. Yet, there has always been a structuralist tradition distinct from (and often in conflict with) the mainstream neo-Darwinian view, in which it is considered important to give explanations of the forms of things in terms of structural principles, e.g., relating the form of a mushroom to the form of a mushroom cloud. One would associate structuralism historically with authors such as D'Arcy Thompson, and in more recent times, with authors such as Goodwin (1994), Kauffman (1993), and Fontana (2002).

Contemporary structuralists often make arguments relating the chances of evolving a particular form to its relative accessibility or density in state-space (e.g., Kauffman 1993, Fontana 2002), and these arguments can be understood as implications of biases in the introduction of variation. To understand how, let us begin with an argument from molecular evolution about the rough correspondence observed between the frequency with which an amino acid is found in proteins, and the number of codons assigned to it in the genetic code (which ranges from 1 to 6). Originally, King (1969) pointed to this correlation as an argument for neutral evolution, but King (1971) later stepped back from this position, arguing that the same correlation could arise under a stochastic view with selective allele fixation, explaining this position as follows:

> Suppose that, at a given time, there are several possible amino acid substitutions that might improve a protein, and among these are changes to serine and to methionine; serine, with its six codons, has roughly six times the probability of becoming fixed in evolution. Once this has occurred, the mutation to methionine may no longer be advantageous. (p. 89)

Amino acids with more codons occupy a greater volume of sequence space. By the same token, they have more mutational arrows pointed at them: they are more likely to be proposed, and therefore more likely to evolve, other things being equal. This is analogous to the argument of Garson, et al. (2003) in "How development may direct evolution," where there is a discrete map between genotypes and phenotypes, and the chance of evolving is a function of the number of genotypes assigned to a phenotype.

The resulting tendency for the frequency of an amino acid (phenotype) to correspond to its number of codons (genotypes) is not caused by natural selection, even if all changes are beneficial: instead, this pattern arises from the way in which the space of possibilities is explored by the mutation process. This is why, in the arguments of Kauffman (1993), the "ensemble properties" whose emergence Kauffman attributes to "self-organization" still arise when differential selection is removed.

A related idea concerns the role of the local accessibility of genotypes with a given alternative phenotype. In the RNA-based models of Fontana (2002), sequences that have the same fold (phenotype) may be represented by a network as in Fig. 6C (after Fontana 2002). An RNA under selection to maintain function may diffuse neutrally within the network for its current fold; the chances of evolving a different fold (phenotype) depend on the accessibility of the new fold over the entire network. For the networks shown in Fig. 6C, a sequence with phenotype $P_0$ is more likely to evolve to $P_2$ than $P_1$, because a random step out of the $P_0$ network (gray) is more likely to yield $P_2$ (green) than $P_1$ (blue).

This is not the same as an effect of the number of genotypes assigned to an alternative phenotype: $P_1$ and $P_2$ both have exactly 13 genotypes (sequences). Instead, it is an effect of the number of alternative genotypes *within the mutational horizon* (the set of 1-mutant neighbors) of the network for the current phenotype.

On this basis, we can see that a key concern of structuralism— understanding the relative chances for the evolution of various forms by considering their distribution and local accessibility in genetic state-space— relies on the action of biases in the introduction of variation. This means that, in ideal cases, we can take advantage of the theory provided above, which allows one to combine biases in the introduction of variants and biases in their fixation within a common framework. This is important because it represents a genuine framework of dual causation that supports quantitative links between causes and effects. By contrast, in the history of evolutionary thought, the dual causation of evolution by internal and external causes is typically addressed by first declaring a fundamental asymmetry by which the dominant cause governs the subordinate cause, or an asymmetry by which one kind of cause is nuanced and quantitative, whereas the other is a matter of yes or no.

## A contrast of theories

One's immediate reaction to the suggestion that there are fundamental but previously unrecognized implications of biases in mutation may be one of bafflement. Aren't these merely straightforward implications of discrete events of mutation— a kind of event that scientists have known for over a century? How could such a fundamental idea escape recognition for 80 years after theoreticians began in earnest to explore the evolutionary implications of genetics in the 1920s?

Actually, the history of evolutionary thought provides multiple reasons to believe that, through the latter half of the 20[th] century, a well informed evolutionary biologist would be strongly disposed *against* recognizing or appreciating the concept of biases in the introduction of variation. The suspicion of internal causes in the dominant neo-Darwinian culture ran so deep that every internalist idea, no matter how reasonable, was treated as an appeal to vitalism. As explained below, the hypothesis that variational biases could influence evolution was considered by the founders of theoretical population genetics *and was specifically rejected as impossible*. Until roughly 1969, theoreticians did not construct models of evolutionary change using an introduction process, because the traditional approach assumed that evolution begins with all relevant alleles already present. This gene pool assumption was part of a neo-Darwinian theory by which evolution has no direct dependence on the chance occurrence of new mutations. This combination of views favored a view of causation in terms of mass-action forces, whereas individual events of mutation that introduce novelty were labeled as proximate causes acting at the wrong level to be important in evolution.

Exploring these barriers helps to clarify the significance, for evolutionary thinking, of recognizing biases in the introduction of variation as potentially important causes of evolutionary bias.

## *The "opposing pressures" argument*

In the mid-1980s, evolutionary developmental biologists proposed a reappraisal of the role of development in evolution via the concept of developmental "constraints." Some uses of this term corresponded to the previous idea of "orthogenesis", meaning the generation of variation in a developmentally preferred direction, rather than indiscriminate "ambigenesis" in all directions (Popov 2009). The workshop article on developmental constraints by Maynard Smith and colleagues (1985) became a touchstone, cited over 1000 times today.

What became of this idea? Years later, critics such as Reeve and Sherman (1993) could point to this seminal paper and complain that there was "no delineation of a mechanism", only a description of the idea of developmental biases in variation. That is, Maynard Smith, et al. (1985) did not actually offer a mechanistic cause. They wrote that

> Since the classic work of Fisher (1930) and Haldane (1932) established the weakness of directional mutation as compared to selection, it has been generally held that directional bias in variation will not produce evolutionary change in the face of opposing selection. This position deserves reexamination.

However, the issue was not re-examined until the work of Yampolsky and Stoltzfus (2001).

Fisher and Haldane showed that, starting with a population of A1 individuals, an inequality of forward and backward mutation rates favoring allele A2 will not result in fixation of A2 if this is opposed by selection. Instead, given that mutation rates are small in comparison to selection coefficients, a deleterious allele will persist at a low level reflecting a balance between the opposing pressures of mutation and selection. In the simplest haploid case, the "mutation-selection balance" is approximated by $f = u / s$, where $u$ is the forward rate of mutation to A2 (the backward rate hardly matters) and $s$ is the selective disadvantage of A2.

This kind of equation makes it easy to depict mutation and selection as opposing forces, with selection pushing down the frequency of A2 with magnitude s, and mutation pushing the frequency up with magnitude $u$. Given that selection coefficients on the order of $10^{-2}$ or $10^{-3}$ were thought to be common, whereas mutation rates were never so large, a wide-ranging conclusion seemed warranted:

> For mutations to dominate the trend of evolution it is thus necessary to postulate mutation rates immensely greater than those which are known to occur, and of an order of magnitude which, in general, would be incompatible with particulate inheritance . . . The whole group of theories which ascribe to hypothetical physiological mechanisms, controlling the occurrence of mutations, a power of directing the course of evolution, must be set aside. (ch. 1 of Fisher 1930b).

By treating mutation and selection as opposing pressures on allele frequencies, Fisher and Haldane reduced the vexing issue of internal orienting factors to *a simple matter of size*. No difficult empirical analysis of evolutionary history was necessary to evaluate different hypotheses for the causes of trends. No experiments were needed. Instead, the entire issue could be solved by extrapolating from a few simple facts and an equation. The argument was breathtaking in its scope and simplicity: population genetics tells us that, because mutation rates are small, internal orienting factors acting via mutation are ruled out, *completely*.

Provine (1978) identifies this as one of the vital contributions of population genetics theory to the success of the OMS (original Modern Synthesis), on the grounds that it appeared to eliminate the rival theories of orthogenesis and Mendelian-mutationism. Indeed, the architects of the OMS repeatedly denied mutational causes based on this argument (e.g., p. 56 of Huxley 1942, p. 114 of Simpson 1953, p. 355 of Mayr 1960, p. 7 to 9 of Mayr 1959, p. 391 of Ford 1971).

This is how the "opposing pressures" argument, to the effect that mutation is a weak force unable to overcome selection or influence the direction of evolution, emerged from Fisher (1930b, a) and Haldane (1932, 1933), as well as Wright (1931). More recent sources also cite this argument directly (e.g., p. 510 of Gould 2002), or draw on the argument that mutation is a weak force (e.g., see Section 5.3 of Freeman and Herron 1998, pp. 105-118 of Sober 1987, or p. 282 of Maynard Smith et al. 1985). This even happens in the literature of molecular evolution (e.g., p. 54 of Li, Luo, and Wu 1985). An interesting example is from a white paper on evolution, science and society endorsed by various professional societies (Futuyma 1998). One section discusses accomplishments of mathematical population genetics, claiming "it is possible to say confidently that natural selection exerts so much stronger a force than mutation on many phenotypic characters that the direction and rate of evolution is ordinarily driven by selection even though mutation is ultimately necessary for any evolution to occur."

## *Population genetics without the introduction process*

For Fisher and Haldane to reject a role for mutation bias without considering its effect on the introduction process may seem absurd to us today. This absurdity is a measure of how far evolutionary thinking has drifted from the OMS. As noted above, the architects of the OMS were committed to a buffet view in which all the variants relevant to the outcome of evolution are present initially. The nature of the OMS view is explored in the next section: our focus in this section is on its impact on modeling.

When the alleles relevant to the outcome of evolution are present initially, the influence of mutation is merely to shift the relative amounts of the alleles. These effects are minor and typically negligible compared to the frequency-shifting effects of selection and even drift, because mutation rates are so small. That is, mutation is a weak pressure because mutation rates are small. As Lewontin (1974) noted, "There is virtually no qualitative or gross quantitative conclusion about the genetic structure of populations in deterministic theory that is sensitive to small values of migration, or any that depends on mutation rates" (p. 267).

As a result, mutation was often simply left out of models. Classic works such as Lewontin and Kojima (1960) don't include mutation because its effects are trivial. The treatment of the mathematical foundations of population genetics by Edwards (1977) includes hundreds of equations, none with a term for mutation: the word "mutation" occurs only on page 3, in the sentence "All genes will be assumed stable, and mutation will not be taken into account."

Thus, there was no theory relating the rate of evolution to the rate of mutational introduction. The need for such a theory became obvious in the 1960s, when protein sequence comparisons suggested that evolution was a recurring process of mutation and fixation. It was in this context that origin-fixation models emerged in 1969 (McCandlish and Stoltzfus 2014).

However, for decades after their first appearance, origin-fixation models remained associated with molecular evolution, and were used primarily to address the fate of neutral or slightly deleterious mutations (McCandlish and Stoltzfus 2014). Outside of molecular evolution, theoretical population genetics still relied on the gene-pool assumption. In the 1990s, theoreticians began to notice this restriction, e.g., Yedid and Bell (2002) write

> In the short term, natural selection merely sorts the variation already present in a population, whereas in the longer term genotypes quite different from any that were initially present evolve through the cumulation of new mutations. The first process is described by the mathematical theory of population genetics. However, this theory begins by defining a fixed set of genotypes and cannot provide a satisfactory analysis of the second process because it does not permit any genuinely new type to arise.

Likewise, Hartl and Taubes (1998) write

> Almost every theoretical model in population genetics can be classified into one of two major types. In one type of model, mutations with stipulated selective effects are assumed to be present in the population as an initial condition . . . The second major type of models does allow mutations to occur at random intervals of time, but the mutations are assumed to be selectively neutral or nearly neutral.

Eshel and Feldman (2001) make a similar distinction:

> We call short-term evolution the process by which natural selection, combined with reproduction (including recombination in the multilocus context), changes the relative frequencies among a fixed set of genotypes, resulting in a stable equilibrium, a cycle, or even chaotic behavior. Long-term evolution is the process of trial and error whereby the mutations that occur are tested, and if successful, invade the population, renewing the process of short-term evolution toward a new stable equilibrium, cycle, or state of chaos.(p. 182)

They argue that

> Since the time of Fisher, an implicit working assumption in the quantitative study of evolutionary dynamics is that qualitative laws governing long-term evolution can be extrapolated from results obtained for the short-term process. We maintain that this extrapolation is not accurate. The two processes are qualitatively different from each other. (p. 163)

Thus, due to the way that prevailing views shaped approaches to modeling, theoreticians were not prepared to recognize a role for biases in the introduction of variation. A half-century after theoretical population genetics emerged as a discipline, the introduction process began to appear as an unnamed technical feature of certain types of models, though not as a feature of contending theories (which were focused instead on selection, drift, recombination, sex, and so on). It took several more decades for the argument to emerge that recognizing the introduction process is not a minor technical detail, but a major theoretical innovation that challenges previous thinking about how evolution works, and opens new avenues for theoretical and empirical research.

## *Structure and implications of the OMS*

In the early years of the 20th century, Johannsen's experiments with beans showed that selection can be effective at sorting out existing varieties, but does not create new types from masses of environmental fluctuations. That experiment spelled the end of Darwinism among those who embraced genetics (Gayon 1998). In the mutationist view that emerged among early geneticists such as Morgan (1916), "Evolution has taken place by the incorporation into the race of those mutations that are beneficial to the life and reproduction of the organism" (p. 194).

Given this view, one might imagine that the introduction of a new mutation would be a key event that provides initiative for evolutionary change, and thus influences dynamics as well as direction. Following Shull (1936), one might suppose that "If mutations are the material of evolution, as geneticists are convinced they are, it is obvious that evolution may be directed in two general ways: (1) by the occurrence of mutations of certain types, not of others, and (2) by the differential survival of these mutations or their differential spread through the population" (p. 122), and we might suppose that a new allele "produced twice by mutation has twice as good a chance to survive as if produced only once" (p. 140).

We might think this way today, but the architects of the OMS emphatically did not. In their view, Johannsen's experiments showed nothing. Instead, the true nature of evolution was revealed in Castle's experiments with the hooded rat, because unlike Johannsen, "Castle had been able to produce new types by selection" (p. 114 of Provine 1971). Castle was able to shift coat-color from mottled to nearly all black, or nearly all white, in less than 20 generations of selection, not enough time for new mutations to play any important role, i.e., *selection can create new types without mutation*.

The genetic interpretation of this result was that selection simultaneously shifts the frequencies of available alleles at many loci, leveraging recombination to combine many small effects in one direction (Provine 1971). Thus recombination (not mutation) is the proximate source of new genetic variation every generation. Evolution, rather than being a process of the mutational introduction and reproductive sorting of variation, is envisioned as a process of shifting allele frequencies in the gene pool. Even though this mode of change requires abundant standing variation, it prevails in nature (they argued), because natural populations have a "gene pool" that "maintains" variation. Thus, the maintenance of variation in the gene pool, logically necessarily to prop up the Castle experiment as the paradigm of evolution, became a major theme, what Gillespie (1998) called the "Great Obsession" of population genetics.

Thus, in the OMS, the term "gene pool" is not merely descriptive, but evokes the theory that natural populations *maintain* abundant genetic variation, e.g., Stebbins (1966) writes that "a large 'gene pool' of genetically controlled variation exists at all times" (p. 12). This theory, proposed by Chetverikov and popularized by Dobzhansky, holds that various features of genetics and population genetics— including recessivity, chromosome assortment, crossing over, sexual mixis, frequency-dependent selection, and heterosis— come together to create a dynamic in which variation is soaked up like a "sponge" and "maintained." In the OMS, this gene-pool dynamic ensures that evolution is always a multi-factorial process in which selection never waits for a new mutation, but can shift the population to a completely new state based on readily available variation.

This gene-pool theory was used to argue against the early geneticists in regard to (1) the source of initiative in evolution (mutation or selection), and (2) rates of evolution, which do not depend on mutation rates, e.g., Stebbins (1959) writes

> Second, mutation neither directs evolution, as the early mutationists believed, nor even serves as the immediate source of variability on which selection may act. It is, rather, a

> reserve or potential source of variability which serves to replenish the gene pool as it becomes depleted through the action of selection . . . The factual evidence in support of these postulates, drawn from a wide variety of animals and plants, is now so extensive and firmly based upon observation and experiment that we who are familiar with it cannot imagine the appearance of new facts which will either overthrow any of them or seriously limit their validity. (p. 305)

Stebbins shows an irreversible commitment to denying any direct role of mutation in evolution. An episode of evolution begins when the environment changes, bringing on selection of an abundance of small-effect variation in the gene pool, so that in the words of Mayr (1963), "mutation merely supplies the gene pool with genetic variation; it is selection that induces evolutionary change" (p. 613).  As Mayr (1963) explains at greater length,

> In the early days of genetics it was believed that evolutionary trends are directed by mutation, or, as Dobzhansky (1959) recently phrased this view, 'that evolution is due to occasional lucky mutants which happen to be useful rather than harmful.' In contrast, it is held by contemporary geneticists that mutation pressure as such is of small immediate evolutionary consequence in sexual organisms, in view of the relatively far greater contribution of recombination and gene flow to the production of new genotypes and of the overwhelming role of selection in determining the change in the genetic composition of populations from generation to generation (p. 101).

The gene-pool theory has a particular implication about rates of evolution. If an event of mutation is never the token cause that initiates evolutionary change, then the rate of evolution will not depend in any strong way on the rate of mutation, e.g., Stebbins (1966) writes:

> mutations are rarely if ever the direct source of variation upon which evolutionary change is based. Instead, they replenish the supply of variability in the gene pool which is constantly being reduced by selective elimination of unfavorable variants. Because in any one generation the amount of variation contributed to a population by mutation is tiny compared to that brought about by recombination of pre-existing genetic differences, even a doubling or trebling of the mutation rate will have very little effect upon the amount of genetic variability available to the action of natural selection. *Consequently, we should not expect to find any relationship between rate of mutation and rate of evolution. There is no evidence that such a relationship exists*. [my emphasis] (p. 29).

Likewise, in their 1970s textbook, Dobzhansky, et al (1977) write:

> The large number of variants arising in each generation by mutation represents only a small fraction of the total amount of genetic variability present in natural populations. … It follows that rates of evolution are not likely to be closely correlated with rates of mutation . . . Even if mutation rates would increase by a factor of 10, newly induced mutations would represent only a very small fraction of the variation present at any one time in populations of outcrossing, sexually reproducing organisms (p. 72).

## *The forces theory*

The structure of the OMS encourages a particular conception of evolutionary causes as frequency-shifting forces.  Evolution at the phenotypic level is envisioned as a smooth shift to a new adaptive optimum, as in Fig. 7A, with a rate and direction that depend entirely on the rate and direction of the environmental changes to which selection is responding. At the genetic level, evolution occurs by shifting the frequencies of alleles already present in the population. This is a multifactorial process, i.e., an episode of adaptation is a shift in frequencies of alleles at many loci simultaneously (Fig. 7B).

If so, then evolution can be understood as a process that takes place in the *topological interior* of an allele-frequency space (e.g., the interior of the cube in Fig. 7C, but not its edges or surfaces); evolution is more specifically a selection-driven shift from a previous optimum to a new multi-locus optimum (e.g., from $t_1$ to $t_2$ in Fig. 7C). Forces are conceived as processes that can displace an evolving system in this interior space. Thus the forces jointly determine the rate and direction of movement. As noted by Fisher and Haldane, the forces are not all equal in strength. Selection is typically the most powerful, and mutation the weakest. For alleles with intermediate frequencies, even drift will have a larger impact than mutation (except in very large populations).

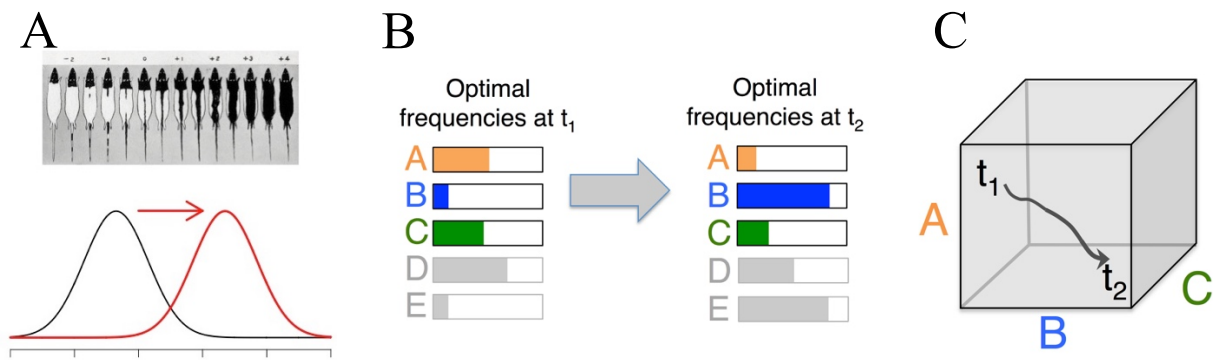

Figure 7. The forces view of the OMS. In the OMS, evolution happens when the environment changes, and (A) selection shifts the phenotypic value, sometimes shifting it so far as to result in a new type. (B) At the genetic level, this shift is a multi-factorial process, dependent on small-effect variation at many loci. Change is not based on mutation-fixation events, but quantitative shifts in frequency of alleles maintained in the gene pool. Therefore, (C) we can understand the process as taking place in the topological interior of an allele-frequency space (i.e., the interior of the cube, but not its edges or surfaces). In this interior space, the trajectory of an evolving system can be understood with the theory of forces. However, the forces theory becomes inadequate when movement occurs outside of this interior space.

For instance, consider just 2 of the dimensions in Fig. 7C. We will say that the position from 0 (origin) to 1 (top) in the A dimension is the frequency of allele <u>A2</u>, as distinct from <u>A1</u>, and in the B dimension, likewise for <u>B2</u> and <u>B1</u>. We can imagine a small shift from the center of this area {0.500, 0.500} to {0.506, 0.506}. In a population of 1000, that would mean adding 6 more <u>A2</u> and 6 more <u>B2</u> individuals. We can imagine such a shift happening by drift, selection, or mutation, although in practice, a shift of ~ 1 % is too much for mutation to accomplish in a generation: it is indeed a weak force.

We could change the details of this example in many ways and the statement would still be true, and this is the whole point of the force analogy. A force is a force because it has a generic ability to shift a frequency $f$ to $f + \delta$. The forces all have a common currency of causation, which is allele-frequency shifts. By way of this common currency, they have a kind of comparability and, in ideal cases, are combinable, i.e., we can combine the shift caused by one force with the shift caused by another. The mutation-selection balance equation would be a simple example of combining two forces using their common currency of causation.

For our purposes, the most important feature of the forces theory is that it fails to include the novelty-introducing role of mutation, which renders it inadequate once we step outside the buffet regime assumed by the OMS. The unique novelty-introducing role of mutation— but not selection, drift, or recombination— is represented by its ability to move the population from $x = 0$ to $x = 1 / N$ (or from $x = 1$ to $x = 1 - 1 / N$). The shift from {0.500, 0.500} to {0.506, 0.506} is mathematically identical to the shift from {0.000, 0.000} to {0.006, 0.006}, but it is not evolutionarily identical: the latter shift absolutely requires the involvement of mutation. Mutation allows the system to jump off of an axis into the interior where the forces of selection and drift operate. Yet, this role is not covered by the "forces" formalism.

That is, the introduction process draws on a different currency of causation inaccessible to selection or drift. This is why the forces theory fails to provide correct reasoning for the behavior of the Yampolsky-Stoltzfus model in the sushi conveyor regime, or even in regimes that are close to it. We described these failures earlier: neither increasing the magnitude of selection coefficients (which increases the force of selection), nor decreasing the magnitude of mutation rates (which decreases the force of mutation), causes selective preferences to overpower mutational ones.

The limitations of the forces view are also revealed in how scientists construct hypotheses and explanations. In molecular evolution, we are regularly faced with comparative evidence suggesting that molecular features, particularly aggregate features like codon usage or proteome composition, are systematically shaped by mutation biases. By the logic of opposing forces, this must mean that the force of selection is effectively absent. Accordingly, for decades, molecular evolutionists habitually treated mutational explanations as references to neutral models (e.g., Sueoka 1988, Gillespie 1991, Gu, Hewett-Emmett, and Li 1998, Knight, Freeland, and Landweber 2001, Lafay et al. 1999, Wolfe 1991).

This interpretation literally suggests that the pressure of recurrent mutation is driving alleles to fixation in the absence of opposing selection. Although this is the correct way to apply the forces theory, the resulting model is not very reasonable, because fixation by recurrent mutation is too slow. The more likely basis for mutation-biased neutral evolution would be that fixation happens by drift, while the mutational bias resides in the introduction process. That is, once we recognize two kinds of causes with two different currencies— introduction and fixation—, this immediately suggests a different interpretation in which the mutation bias is applied only to the introduction process. From this point, it is a small step to consider that the fixation process might be either drift or selection, leading to the novel prediction of mutation-biased adaptation.

## Conclusion

The notion that a fundamental cause-effect relationship could emerge 80 years after the origins of theoretical population genetics must be considered surprising, as suggested by the epigraph at the head of this chapter (Charlesworth 1996). One eminent theoretician argued that explaining the nature of the results of Yampolsky and Stoltzfus (2001) was not a worthwhile activity because we have "matured past the phase when discussions of basic principles are useful for professionals. We all believe in mutation, selection, and drift, I hope." (Stoltzfus 2012). Lynch (2007) cites Yampolsky and Stoltzfus (2001) along with a handful of other sources going back to *Charles Darwin*, explaining that "the notion that mutation pressure can be a driving force in evolution is not new."

A theory is new if scientists have not considered it before, either because it was never introduced, or because it was introduced and then ignored. The question of whether some set of past statements about mutation or variation constitutes a prior description of the kind of cause-effect relationship described here is a matter of applying the duck test: if it looks like a duck, swims like a duck, and quacks like a duck, it's a duck. Either the distinctive implications of biases in the introduction process have been previously shown to arise from a consideration of basic principles of population genetics, or they have not.

One's disbelief at the belated discovery of this kind of cause-effect relationship must begin to fade rapidly when the historical development of evolutionary thought is considered. In fact, (1) the founders of theoretical population genetics considered the possible role of biases in variation as an influence on the course of evolution, (2) they reached the conclusion that such an effect was impossible, and (3) this conclusion became a foundation of the OMS, cited until contemporary times. A role for internal tendencies of variation was rejected by leading thinkers *precisely on the grounds of lacking a population-genetic mechanism*. Indeed, the idea was ridiculed, e.g., as when Simpson (1967) argued that selection is the only plausible cause of directional changes in the fossil record, dismissing "the vagueness of inherent tendencies, vital urges, or cosmic goals, without known mechanism" (p. 159).

Decades of results from molecular studies of evolution have conditioned us to accept the mutationist idea that the timing and character of evolutionary change may depend on the timing and character of mutations. Because of this, scientists today may find the evolutionary role of biases in the introduction process to be intuitively obvious. However, the foundations of contemporary evolutionary thought were laid by scientists who explicitly rejected mutationist thinking as an apostasy, and who were committed to the doctrine that variation plays only the role of a material cause, a source of substance only, never a source of form or direction.

From this context, we can begin to see that evolutionary reasoning about causes and explanations is not merely a matter of mathematics, and is not merely a matter of following obvious implications of basic principles that everyone accepts.

Other factors are at work to make the recognition and appreciation of causes non-obvious. We have addressed two of them already. One factor is the role of high-level conjectural theories asserting that evolution works in some ways and not in others, and in particular, the role of neo-Darwinism in shaping the OMS theory of evolutionary genetics. To understand evolutionary reasoning requires, not merely a knowledge of mathematics, but an awareness of the substantive conjectural theories that have shaped prevailing mathematical approaches to evolution, e.g., the gene-pool theory. One cannot understand the strengths and weaknesses of contemporary evolutionary thinking if one does not understand how it relies on the OMS conjecture that evolution operates in the buffet regime of abundant pre-existing variation (Stoltzfus 2017).

A second factor, which requires more explication, is the role of verbal theories of causation and explanation. Even if we attempt to set aside the OMS and other broad conjectures, we must recognize that the conceptual tools we use to negotiate issues of causation and explanation include verbal theories that draw on metaphors and analogies (e.g., pressures, raw materials). Mathematics itself is not a language of causation, but a notation for relating quantities, along with a set of rules for manipulating these relations. The "equals" sign has no direction, thus it cannot possibly represent an arrow of causation. There is no necessary causal translation of $a = bc$, or the mathematically equivalent statements $b = a / c$ or $a / b = c$.

Thus, the rules of mathematics do not compel any particular interpretation of the equations of population genetics. For example, little in the breeder's equation ($\Delta z = h^2 S$) differentiates variation and selection: they are 2 factors multiplied together to yield a product. If we choose to call the product "the response to selection", rather than something else, it is because we have situated this equation within some conceptual framework that extends beyond mathematics. Indeed, it is natural to refer to the left-hand side as "the response to selection" because, in practice, selection has nearly always been the experimentally manipulated variable, even though it is possible in principle to manipulate the heritable component of available variation in a population, in which case we might call the left-hand side "the response to variation."

After Lande and Arnold (1983) generalized the breeder's equation to multiple dimensions as $\Delta Z = G\beta$ (where $\beta$ is a vector of selection differentials, $G$ is a matrix of variances and covariances quantifying standing variation, and $\Delta Z$ is a vector of expected changes in trait values), quantitative geneticists began to understand that, in the words of Steppan, et al. (2002), "Together with natural selection (the adaptive landscape) it [the $G$ matrix] determines the direction and rate of evolution." This verbal theory of dual causation, in which the *rate and direction* of evolution are *jointly* attributed to selection and standing variation, was novel: it does not correspond to the previous verbal theories in which the rate and direction of evolution *are governed by selection* (while variation merely supplies raw materials), even though quantitative genetics is the branch of mathematical theory that most closely follows neo-Darwinian assumptions.

That is, Darwin's verbal theory led to Fisher's mathematical theory, and then further mathematical developments along with empirical results (e.g., Schluter 1996, McGuigan 2006) led to conflicts with the original verbal theory, resulting in a new verbal theory that aligns better with mathematical models and empirical results.

This kind of reform, in which verbal theories of causation are replaced or modified so as to achieve better alignment with mathematical models and empirical results, has not yet occurred in regard to the introduction process, and particularly in regard to quantitative biases in the generation of variants in discrete spaces (which are not the same thing as asymmetries in standing variation of continuous traits represented by the $G$ matrix). The architects of the OMS were explicit that events of mutation are not evolutionary causes, but proximate causes that act at the wrong level to be evolutionary causes. The emergence of origin-fixation models in 1969 posed a threat to this view, yet the threat did not materialize: it did not induce a shift in conceptualizations of causation. Neither those who explicitly invoked the need to address long-term dynamics dependent on new mutations, nor those who recognized a renaissance in thinking about adaptation, called for explicit recognition of a point process of introduction absent from the forces theory.

The practice of scientific reasoning is not a form of magic. A complete understanding of the evolutionary implications of mutation does not blossom forth in one's mind simply by uttering the word "mutation." Instead, our understanding of causal factors must be constructed. In practice, it is constructed in a context that includes, not only equations and facts, but low-level folk theories of causation (e.g., pressures, raw materials) and high-level conjectures about how causes work together to account for larger phenomena. Recognizing the introduction process as a causal process in evolution, and then recognizing the consequences of biases in the introduction process, makes variation-biased evolution intelligible in a way that led to the prediction of mutation-biased adaptation (now verified), and which yields new insights into the potential role of development and self-organization in evolution.

To summarize, the novelty of scientific theories is not judged by whether they are considered intuitively obvious with benefit of hindsight, or whether they require a complex mathematical derivation. To recognize a causal role of biases in the introduction process is novel because this kind of causation has distinctly unfamiliar behavior and implications. Generic references to mutation pressure, contingency, chance, constraints, and so on, do not pass the duck test. We know that they are not the same thing as references to biases in the introduction process because *the former were never previously understood to exhibit the implications that the latter can be demonstrated to exhibit*. Generations of theoreticians who invoked the term "mutation" failed to comprehend the implications of biases in the introduction process: again, such comprehension does not emerge magically by uttering the word "mutation," but requires a modeling framework and a conceptualization of causation.

Mutational and developmental biases in the introduction of variation appear to be important determinants of the outcome of evolution, representing a novel kind of causation absent from the Modern Synthesis. The extent of their importance, and particularly their importance as predictable factors acting at the level of phenotypic evolution, is a major unresolved issue in evolutionary biology.

## Acknowledgements

The author thanks David McCandlish, Yogi Jaeger and an anonymous reviewer for helpful comments. The identification of any specific commercial products is for the purpose of specifying a protocol, and does not imply a recommendation or endorsement by the National Institute of Standards and Technology. Figure 1A is a photo by John Johnston (original available via CC-BY). In Figure 3, the sushi conveyor photo is by NipponBill (original) and the salad buffet photo is by Heider Ribeiro (original), both available via CC-BY-SA license.